\documentstyle[aps,epsf]{revtex}  
\newcommand{\et}{{\rm E}_{\scriptscriptstyle\rm T}}       
\newcommand{\pt}{{\rm P}_{\scriptscriptstyle\rm T}}

\newcommand{\MET}{\mbox{$\protect \raisebox{.3ex}{$\not$}\et$}}
\newcommand{\met}{\mbox{$\protect \raisebox{.3ex}{$\not$}\et$}}

\begin{document}        

\baselineskip 14pt
\title{Standard Model Higgs and Top Mass Measurements at the Tevatron}
\author{W.-M. Yao \\ For CDF and D0 Collaboration}
\address{Lawrence Berkeley National Laboratory, One Cyclotron Road,
Berkeley, CA 94720, USA\\E-mail: wmyao@lbl.gov}
%
\maketitle              

\begin{abstract}        
  A summary of the present Standard Model Higgs search and measurement of top 
quark mass at the Tevatron are presented. The sensitivity of the present 
Higgs search at the Tevatron is limited by statistics to a cross section 
approximately two orders of magnitude higher than the 
predicted cross section for standard model Higgs production. 
With 30 fb$^{-1}$ of integrated luminosity, the Tevatron
offers an unique potential discovery window for the Standard Model Higgs mass 
up to 130 GeV/c$^2$ before LHC era. 
The study of top at the Tevatron has moved from discovery phase to one of 
characterizing its properties. The combined result of top quark
mass is $174.3\pm 5.1$ GeV/c$^2$ ($\delta m_t/m_t < 3\%$).
\end{abstract}   	

\section{Search for Standard Model Higgs at Tevatron} 
  For several decades, the standard model has been remarkably successful in 
explaining and predicting experimental data. However,
 the mechanism of  electroweak
symmetry breaking is  still not known. The most popular mechanism to
induce spontaneous symmetry breaking of a gauge  theory, resulting in the gauge
bosons and fermions acquiring masses, is the  Higgs mechanism~\cite{Higgs},
which predicts the existence of a Higgs particle with unknown mass. 
The current direct search limit~\cite{lep2} for Standard Model Higgs Boson 
at LEP-2 is
$m_h \ge 96$ GeV/c$^2$ at 95\% C.L. LEP-2 will continue their run until 
the end of year 2000, which will allow them to either exclude the Higgs mass 
up to 110 GeV/c$^2$ or have a 5 $\sigma$ Higgs discovery for Higgs mass below
106 GeV/c$^2$. The indirect search via electroweak precision 
tests through radiative correction yields $m_h \le 260 $ GeV/c$^2$ at 95\% C.L.

 At the Tevatron, one of the Higgs production processes more likely to be 
observed is an associated production $V+h$ for $m_h \le 130$ GeV/c$^2$, 
where V=W, Z and $h\rightarrow b\bar b$. For the Higgs mass above 
130 GeV/c$^2$, different search strategies need to be developed since the 
dominant decay mode is no longer $b\bar b$, but $W W^*$~\cite{wwst}. 

\subsection{Search for $Wh\rightarrow l \nu b\bar b$} 
The experimental signature considered is  $ W h $ with $W \rightarrow
e\nu$ or $\mu\nu$, and $h\rightarrow b\bar b$,  giving final states with one
high-$\pt$ lepton, large missing transverse energy ($\met$)
 due to the undetected neutrino and 
two $b$ jets. 
The ability to tag $b$ jets
with high efficiency and a low mistag rate is vital for searching for the
decay of $h\rightarrow b\bar b$. 
CDF uses the secondary vertex (SECVTX) and 
soft-lepton (SLT) $b$-tagging algorithms developed for the 
top quark discovery, while D0 uses a soft-muon $b$-tagging only. 
Both CDF and D0 select the $b$-tagged $W+$ 2 jet events since it is 
expected to contain most of the signal, while $b$-tagged $W+\ge 3$
jet events are  dominated by  $t\bar t$ decays.

\begin{figure}[ht]	
\centerline{\epsfxsize 2.0 truein \epsfbox[50 145 540 620]{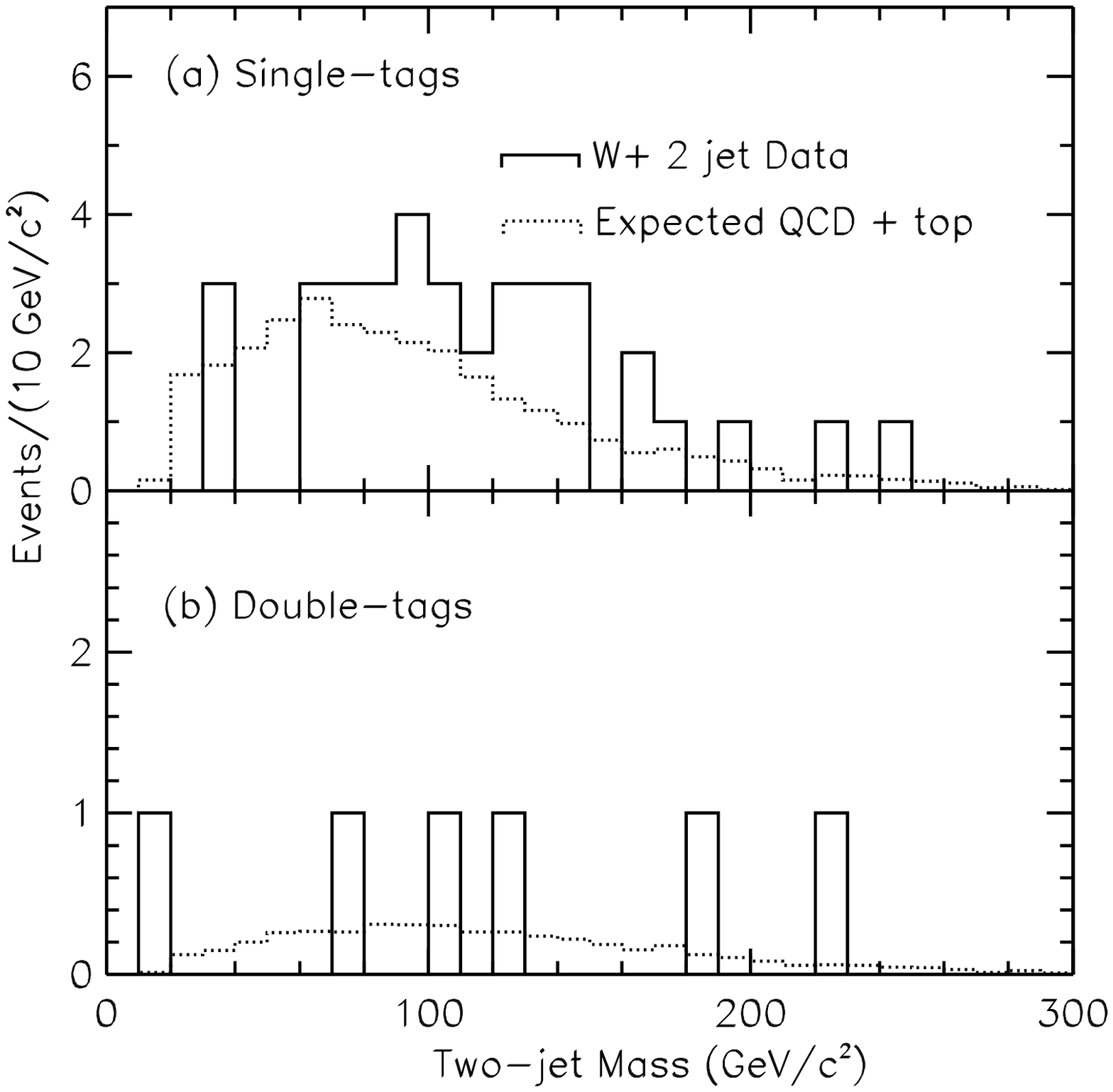}
 \hskip 1.0 cm \epsfxsize 2.0 truein \epsfbox{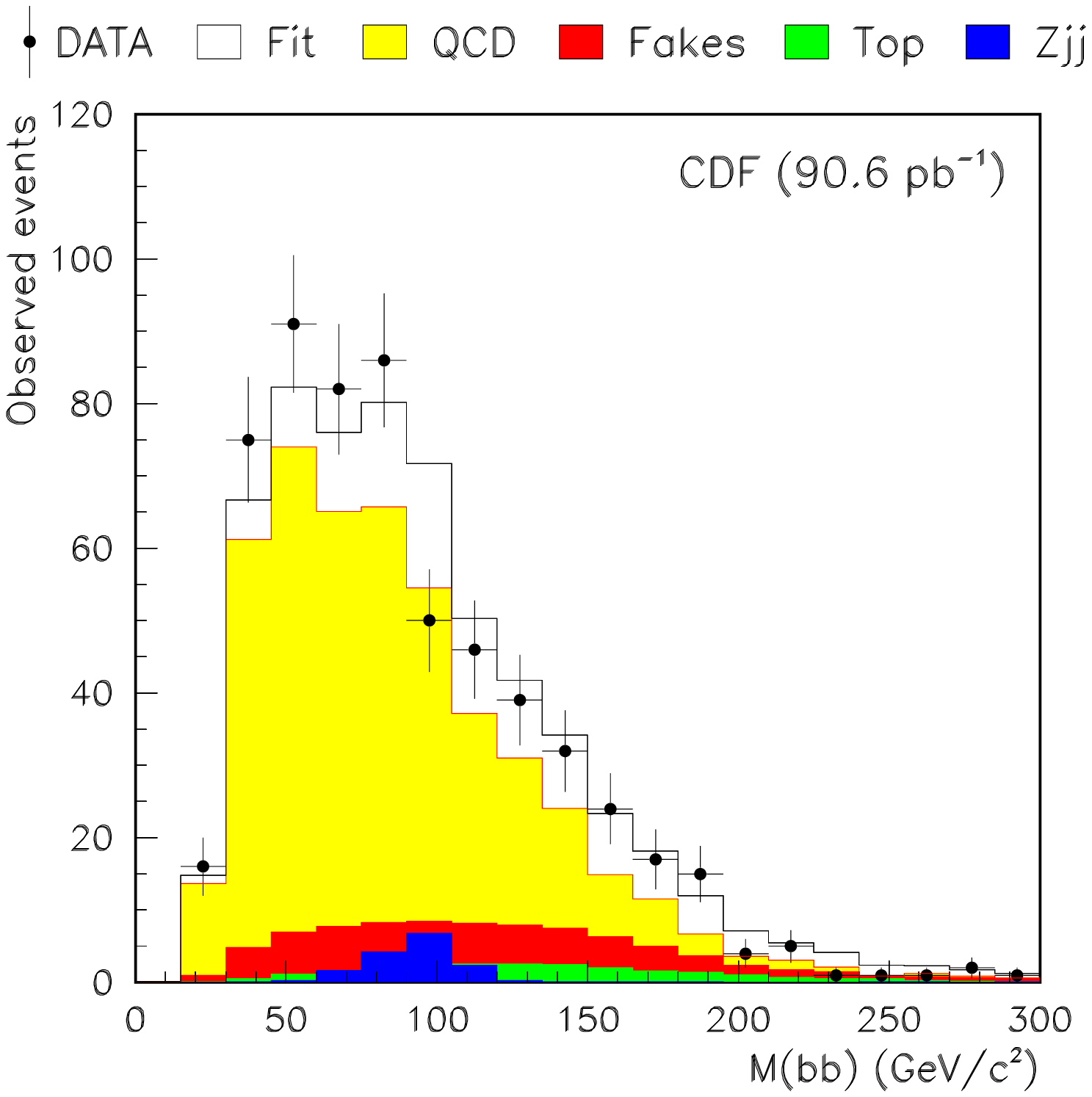} 
 \hskip 1.0 cm \epsfxsize 2.0 truein \epsfbox[0 0 422 441]{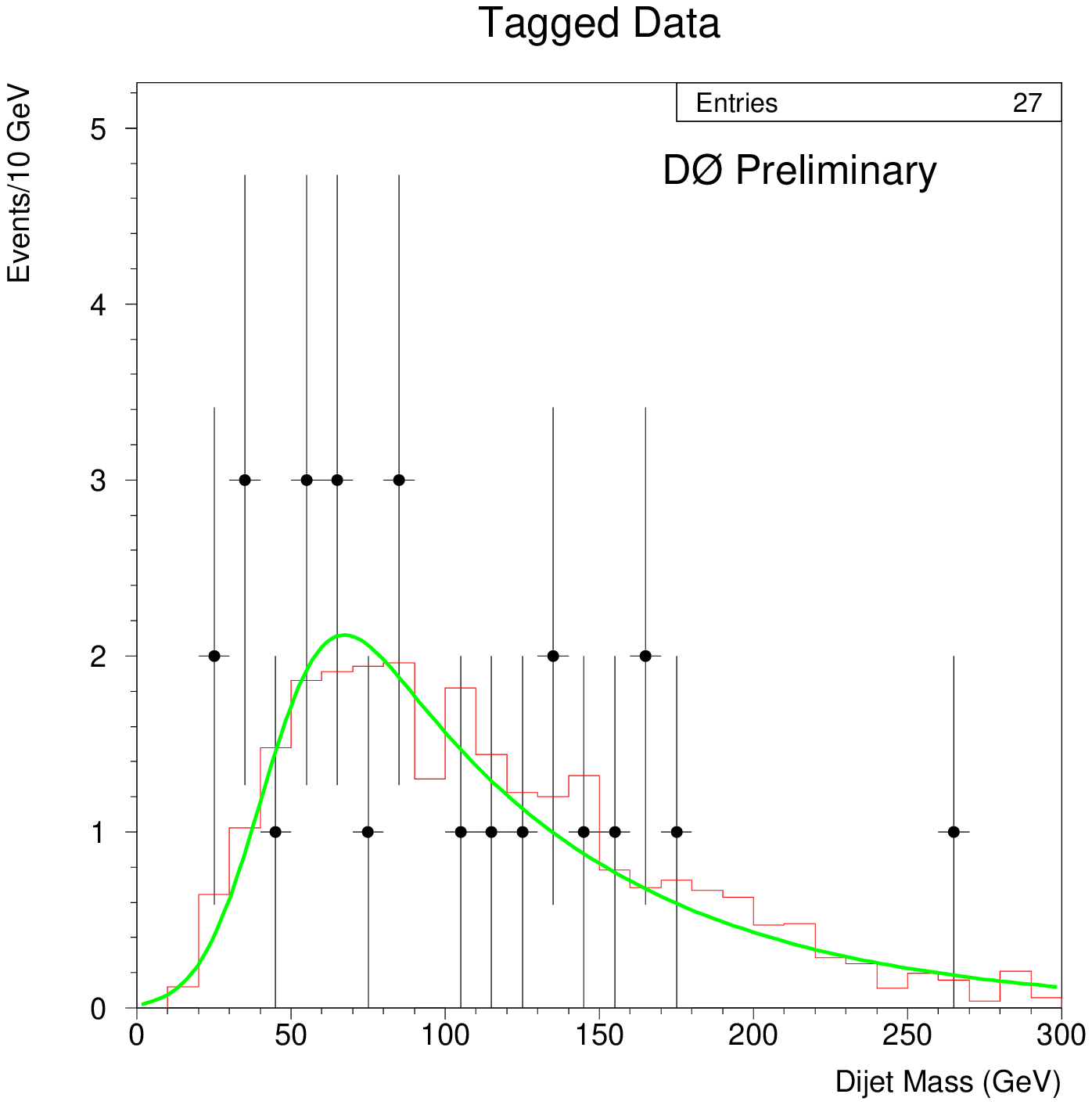}}   
\vskip -.2 cm
\caption[]{
\label{bbmass}
\small The measured two-jet mass distributions along with background 
 in the $Wh\rightarrow l\nu b\bar b$ channel(left) 
 and in $Vh\rightarrow J_1 J_2 b\bar b$ (center) for the CDF(left) and 
 D0(right). }
\end{figure}

In a data sample of $109\pm 9$ pb$^{-1}$, CDF observed 36 events with single
SVX $b$-tag and 6 events with two $b$-tagged jets 
(SVX-SLT and SVX-SVX)~\cite{cdfwh}. The 
expected backgrounds are $30\pm 5$ for single and $3.0\pm 0.6$ for 
double tagged events, which are predominately from $Wb\bar b$, $Wc\bar c$,
mistags and $t\bar t$ decays. The probability that the background
fluctuated upward to the number of observed events is found to correspond 
to one standard deviation. The dijet mass distribution for the single and
double tagged events are shown in Figure~\ref{bbmass}, 
along with the background 
expectation. The single-tag data show a slight excess of events at higher 
two-jet mass but there is  no mass peak as would be expected from the two-body
decay of a new particle. Likelihood fits to the mass distributions yield 
 a 95\% C.L. limit on the 
production cross section times branching ratio, 
shown in Figure~\ref{bblimit}, as a function of Higgs mass.

  D0 also performed similar search in their 100 pb$^{-1}$ 
lepton + jets data~\cite{d0wh}.
 They observed 27 events with at least one $b$-tag, which are in good agreement
with the background expectation of $25.5\pm 3.3$ events. The observed dijet 
mass distribution and corresponding 95\% C.L. limit on the production cross 
section times branching ratio are shown in Figure~\ref{bbmass} and 
\ref{bblimit}.

\subsection{Search for $Vh\rightarrow j_1 j_2 b\bar b$} 

  CDF also searches for Higgs in the four-jets channel of $Vh$ process, where 
V=W,Z decays into two jets~\cite{cdfvh}. The events are required to have 
at least two SVX $b$-tagged jets and the $\pt$ of two $b$-jets greater than 
50 GeV/c$^2$ in order to reduce the large QCD background. In a multi-jet 
trigger sample of $91\pm 7$ pb$^{-1}$, CDF observed 589 candidate events. 
The invariant 
mass of two $b$-tagged jets is shown in Figure~\ref{bbmass}, 
along with the main
backgrounds from QCD, fake tags and $t\bar t$ decay. A likelihood fit is then
applied to the shape of signal plus background, with the normalization of 
signal and QCD free, which yields total 600 background events and consistent
with zero signal events. The corresponding 95\% C.L. limits 
on the production cross section times branching ratio is shown in 
Figure~\ref{bblimit}. It also shows the CDF combined limit obtained from both 
all-hadronic and lepton +jets channels.  

\subsection{Search for $Zh\rightarrow (l^+l^-, \nu\bar \nu)b\bar b$} 

 Several studies~\cite{snowmass} 
in past have indicated that there is significant
potential of observing associated production of Higgs and $Z^0$ boson with
$Z^0h\rightarrow (l^+l^-, \nu \bar \nu)b\bar b$, because of large branching 
rations of $Br(Z^0\rightarrow \nu\bar \nu)=19.2\%$. 
The signature of such decay is relative clean, a resonance of two $b$ tagged 
jets, plus either large missing $Et$ ($\MET>40$ GeV) or dilepton pair mass 
near the $Z^0$ mass, which were triggered experimentally by either 
$\MET>35$ or leptons in the event. 

The $\met$ in the background sample is predominately  due to mismeasured 
jets resulting in the $\met$ direction  being aligned along with the
jet direction while the $\delta \phi_{min}(\met,jet)$ in the 
signal sample is flat.
By requiring large missing $\et$ ($\MET>40$ GeV) and 
$\delta \phi_{min}(\met,jet)>1.0$,
the QCD background can be reduced to minimum, which results in a 
Higgs reach sensitivity similar to the one in 
$Wh\rightarrow l\nu b\bar b$ channel. 

\begin{figure}[ht]	
\centerline{\epsfxsize 2.0 truein \epsfbox{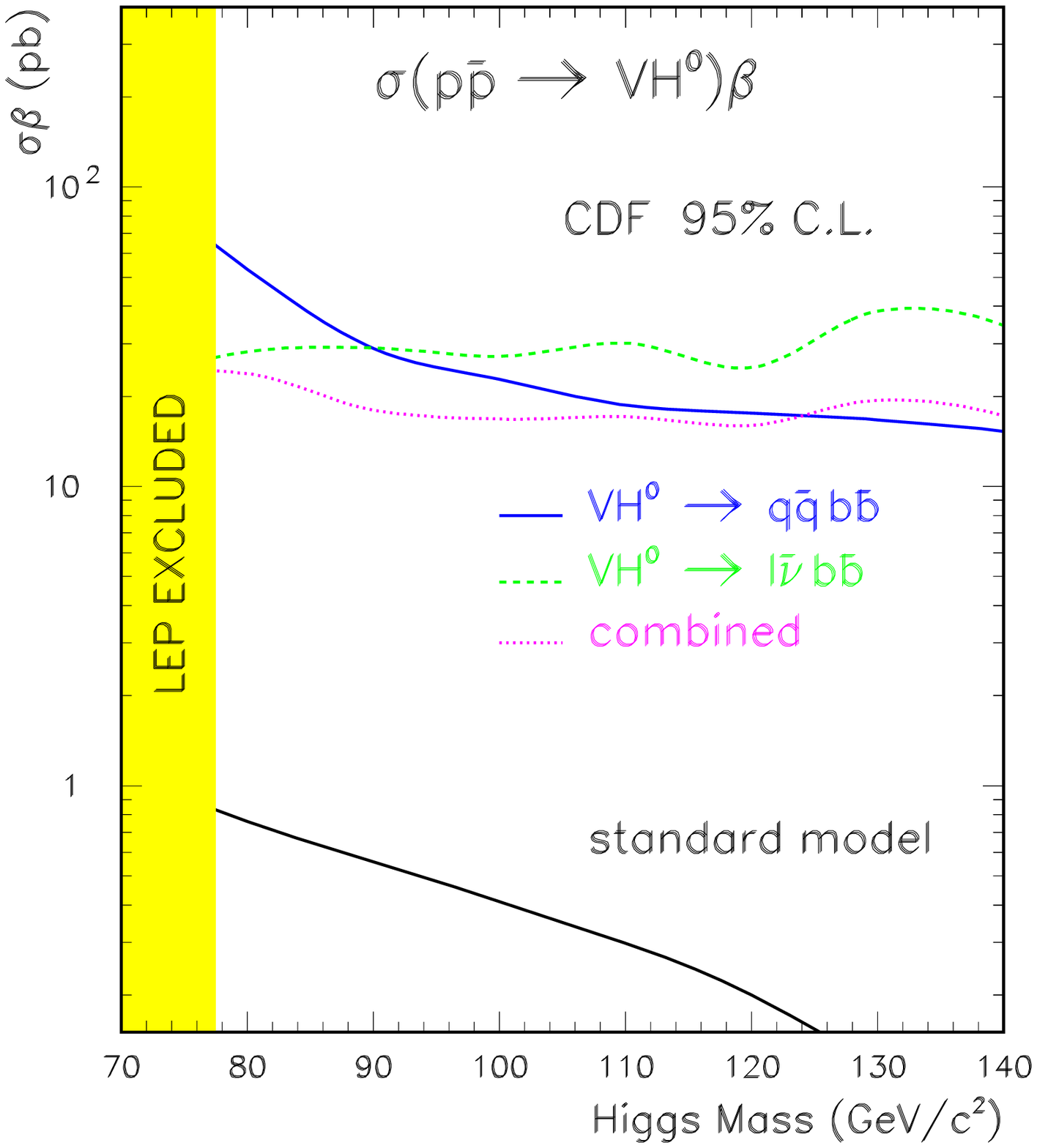}
\hskip 1.0 cm \epsfxsize 3.0 truein \epsfbox[0 0 564 327]{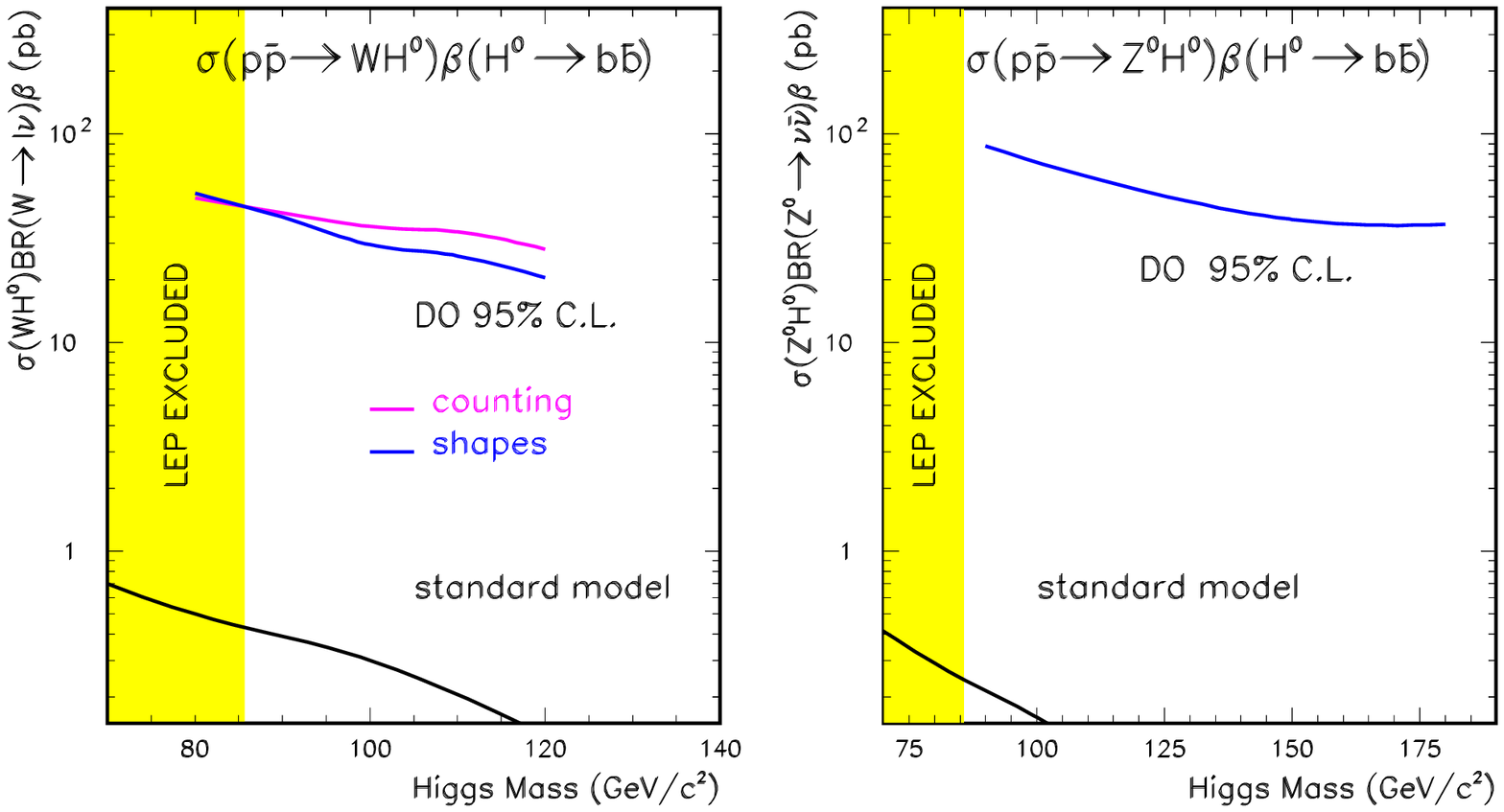}} 
\vskip -.2 cm
\caption[]{
\label{bblimit}
\small  95\% C.L. upper limits on production times branching ratios as function
 of Higgs mass from CDF(left) and D0 (right). } 
\end{figure}

The analysis of CDF search in Run1 is in progress. D0 has 
searched for $ZH\rightarrow \nu\bar \nu b\bar b$ in their Run1 data. 
They observed 2 events in the data, in good agreement with the expected 
background of $2.0\pm 0.7$ events. 
The corresponding 95\% C.L. limits on the production cross section times
branching ratio is shown in Figure~\ref{bblimit}. 

\subsection{The Prospects of Standard Model Higgs Search at Tevatron } 

 During Feb.- Nov. 1998, there was a joint CDF/D0/Theory working group to 
study Higgs searches at Run-II or beyond at the Tevatron 
using more realistic CDF/D0 detector simulation for the signal and 
background estimation 
(see http://fnth37.fnal.gov/higgs.html for more details). 
Since the Tevatron would be the only chance to discover the 
Higgs before LHC, it would be important to establish the realistic 
Higgs discovery potential beyond LEP-2. 
Great progress
has been made in terms of extending the depth and breadth of previous
studies on the Higgs reach~\cite{tev2000}. There is 
no single, golden discovery channel at the Tevatron. Combining all the 
channels and data from both experiment data is crucial. 
The preliminary results 
indicate that the Tevatron should be able to exclude the Higgs 
mass up to 130 GeV/c$^2$ at the 95\% C.L. with 2 fb$^{-1}$ data, or have 
a 5 $\sigma$ discovery of the Higgs mass up to 130 GeV/c$^2$ 
with 30 fb$^{-1}$ data~\cite{hobbs}.

\section{Top Mass Measurements} 
 A precise measurement of the top quark mass is an important ingredient
 in testing the consistency of the standard model with experimental
 data. In addition,  precise $W$ and top mass measurements can
 provide information
 on the mass of the Higgs boson, which is a remnant of the mechanism that
 gives rise to spontaneous electroweak
 symmetry breaking.

   Within the framework of the Standard Model the top quark decays almost 
exclusively into a real W boson and a b quark. The
observed event topology is then determined by the decay modes of the two 
$W$ bosons, which can be classified into three decay channels.
The 
Dilepton Channel has about 5\% of the cases, with both $W$ bosons decay to 
$e\nu$ or $\mu\nu$.
The Lepton + Jets Channel has 30\% of the cases, one $W$ boson decays
 to $e\nu$ or $\mu\nu$, and the other to a $q\bar{q}^{\prime}$ pair. 
The All-Hadronic Channel has $44\%$ of the cases, 
involving the hadronic decay of both $W$ bosons. 

\begin{table}
 \caption{ The observed number of events and expected backgrounds for 
the top decay by the CDF and D0 experiments.}
 \begin{tabular}{|c|c|c|c|c|} 
Channels & CDF Events &  BG & D0 Events & BG \\  \hline
Dilepton & 9 & $2.4\pm 0.5$ & 5 & $1.4\pm 0.4$ \\ \hline
Lep+Jets(SVX) & 34 & $9.2\pm 1.5$ & - & - \\ \hline
Lep+Jets(SLT) & 41 & $24.8\pm 2.4$ & 11 & $2.5\pm 0.5$ \\ \hline 
 Lep+Jets (topological) &- & -& 19 & $8.7\pm 1.7$ \\ \hline  
Alljets & 187 & $142\pm 12$ & 44 & $25.3\pm 3.1$ \\ \hline
$e\nu$ & & & 4 & $1.2\pm 0.4$ \\ \hline
$(e,\mu)\tau$ & 4 & $\approx 2 $ & -&- \\ \hline 
\end{tabular}
\end{table}

  Table~I. shows that CDF and D0 observe a clear excess of events over 
the expected background in the dilepton, lepton + jets, and all-hadronic 
channels. The two experiments taking slightly different approaches in 
defining their events samples, with CDF taking advantage of their 
silicon vertex detector (SVX) and D0 making greater
use of kinematic variables to reduce backgrounds. The present sample of 
top candidates are consistent with Standard Model decays. One of 
CDF and D0 physics goals in Run1 is to determine the top mass as 
accurately as possible using many different channels and techniques. As a
result, the uncertainty on the top quark mass 
has improved from $M_{top} = 174\pm 16$ GeV/c$^2$\cite{prd}, first 
published in 1994 to the present $M_{top}=174.3\pm 5.1$ GeV/c$^2$.

\subsection{Lepton + jets Channel} 
  The advantage of measuring a top quark mass in the Lepton + Jets channel is 
 its relatively larger branching ratio and the ability to full 
 reconstruct the top mass on an event-to-event basis. Both CDF and D0
 select events containing a high $\et$ ($\pt$) single isolated electron (muon) 
 in the central region, large missing transverse energy and at least four 
 jets.

Measurement of the top quark mass begins by fitting each event in the 
sample to the hypotheses of $t\bar t$ production followed by decay in the 
lepton + jets channel ($t\bar t\rightarrow W^+bW^-\bar b \rightarrow 
(l^+\nu b)(q\bar q' \bar b)$). There are twelve distinct ways of assigning the
four leading jets to the four partons $b$, $\bar b$, $q$ and $\bar q'$. In 
addition, there is a quadratic ambiguity in the determination of the 
longitudinal component of the neutrino momentum. This yields up to twenty-four
different configurations for reconstructing an event according to the $t\bar t$
hypothesis with no $b$-tag, 12 configurations for events with a single $b$-tag,
and 4 configurations for events with double $b$-tags. Both CDF and D0 
choose the jet configuration with the lowest $\chi^2$. 

    The precision of the top quark mass measurement is expected to increase 
with the number of observed events, the signal-over-background ratio, and the 
narrowness of the reconstructed-mass distribution. 
In CDF lepton + jets mass analysis~\cite{cdflj},  Monte Carlo studies show 
that an optimum way to partition the sample consists of subdividing the events
into the four statistically independent subsamples: SVX single $b$-tag, 
SVX double $b$-tag, SLT $b$-tag and untagged events. 

\begin{figure}[ht]	
\centerline{\epsfxsize 2.5 truein \epsfbox{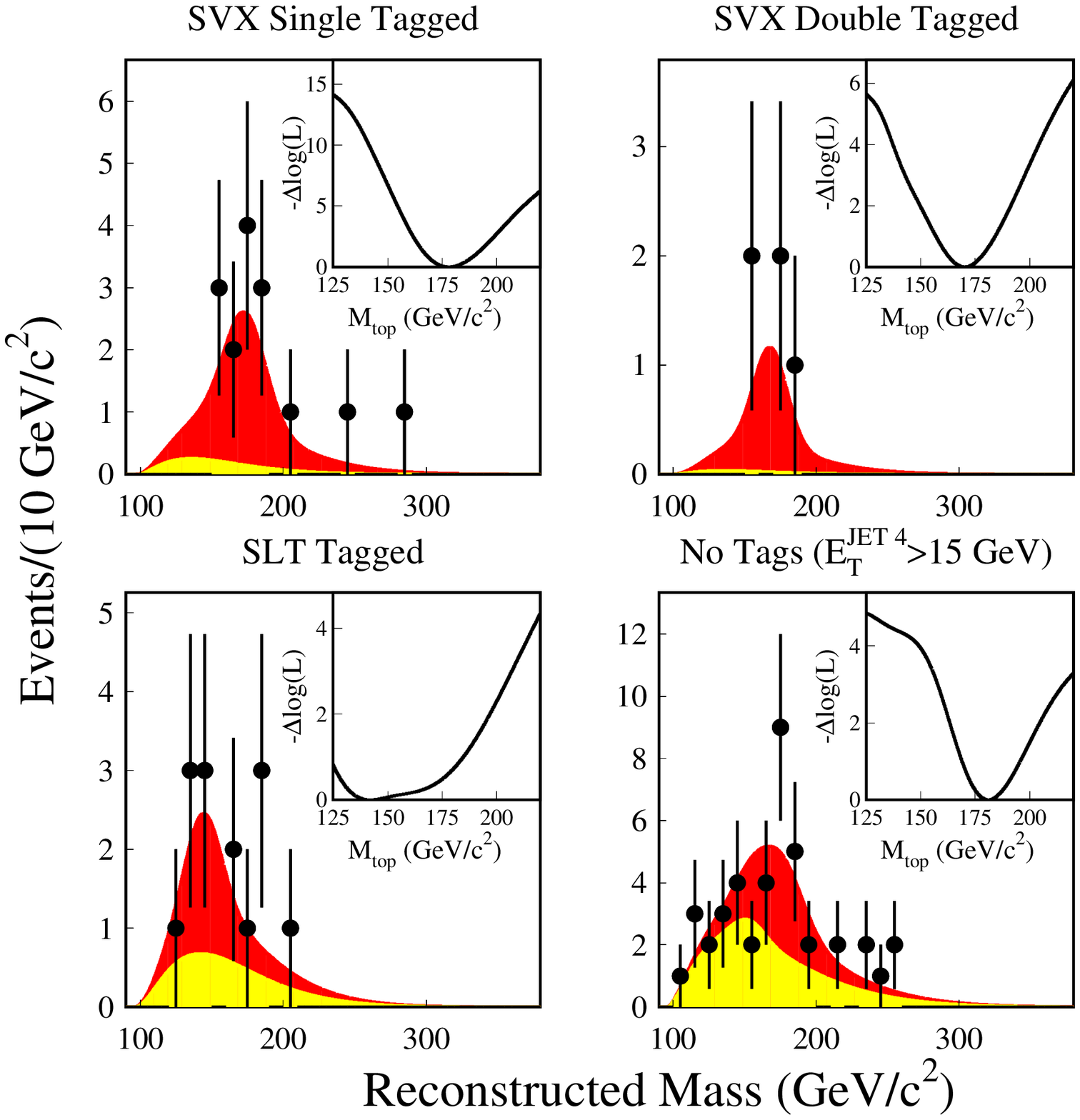 }
\hskip 1.0 cm \epsfxsize 2.5 truein \epsfbox{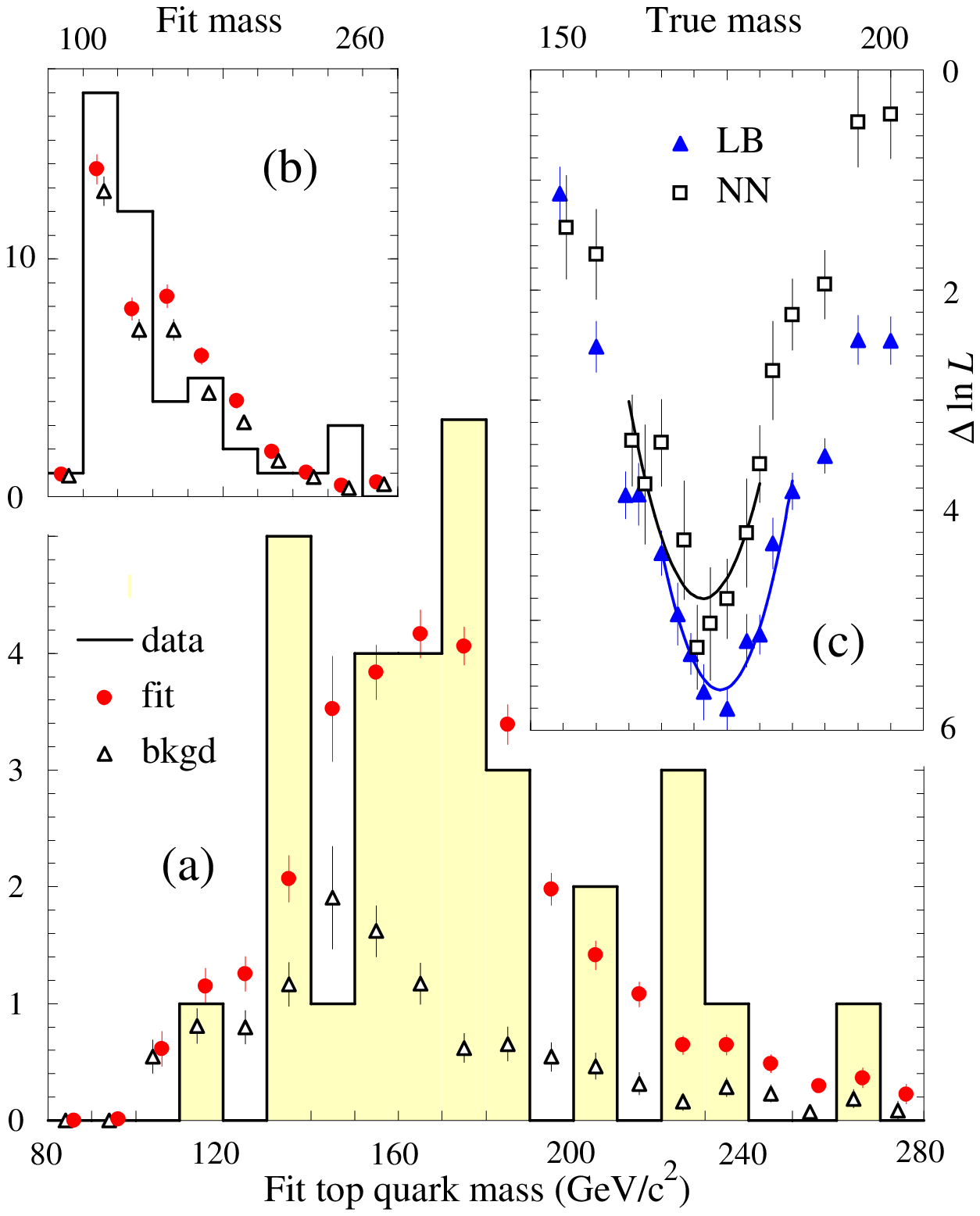 }
}
\caption[]{
\label{cdflj}
\small Left: CDF Lepton + jets mass fits in the SVX single $b$-tag, 
SVX double $b$-tag, SLT $b$-tag and untagged sub-samples, along with 
the background(light shaded) and the fitted top (dark shaded). 
Right: D0 Lepton + jets mass fit for (a) the signal-rich LB sample and 
(b) the background-rich LB sample and (c) the dependence of the likelihoods
on $m_t$.} 
\end{figure}

   The reconstructed-mass distributions of the four subsample 
is plotted in Figure~\ref{cdflj}, from which CDF
measures $m_t = 175.9 \pm 4.8(stat.) \pm 5.3(syst.) $ GeV/c$^2$. 

   The D0 lepton + jets mass analysis uses four kinematic variables
weakly correlated with top mass to separate the top signal from the 
background processes~\cite{d0lj}. 
They were combined into two different multivariate 
discriminant: a weighted likelihood that minimizes correlation with the 
reconstructed mass (LB or ``Low Bias") and a Neural Network output (NN). 
The data are binned 
according to the discriminant value to create top-rich and 
background-rich sub-samples. The subsamples are simultaneously fit to obtain
the top quark mass  $m_t=173.3\pm 5.6(stat.)\pm 5.5 (syst.)$ GeV/c$^2$,
shown in Figure~\ref{cdflj}.

\subsection{Dilepton Channel} 
  CDF recently reported an improved measurement of the top 
quark mass using dilepton events~\cite{cdfdiln}
 originating predominantly from 
$t\bar t \rightarrow W^+b W^- \bar b \rightarrow 
(l^+\nu b)(l^- \bar \nu \bar b)$, where $l = e$ or $\mu$. This measurement 
supersedes their previously reported result in the dilepton 
channel~\cite{cdfdilo}, which was obtained by comparing data with Monte Carlo 
simulation of $t\bar t$ events for two kinematic variables, the $b$-jet 
energies and the invariant masses of the lepton and $b$-jet systems.

 Since the dilepton system is under-constrained due to the two missing 
 neutrinos in the final state, the CDF and D0 dilepton mass 
analyses~\cite{d0dil} proceed by 
hypothesizing a top mass ($m_t$) and solving event kinematics up to four-fold
ambiguity for each of the two lepton-jet parings. Two different 
weighting techniques have been developed. 
 \begin{itemize} 
 \item Neutrino weighting technique assigns a weight to each solution by 
 comparing the predicted and measured missing 
transverse energy (CDF and D0)~\cite{d0dil}. 
 \item Matrix Element weighting technique assigns a weight to each solution by
the parton distribution functions and the matrix element for $W$ boson decay
(D0)~\cite{mew}. 
 \end{itemize} 

\begin{figure}[ht]	
\centerline{\epsfxsize 2. truein \epsfbox[44 136 553 630]{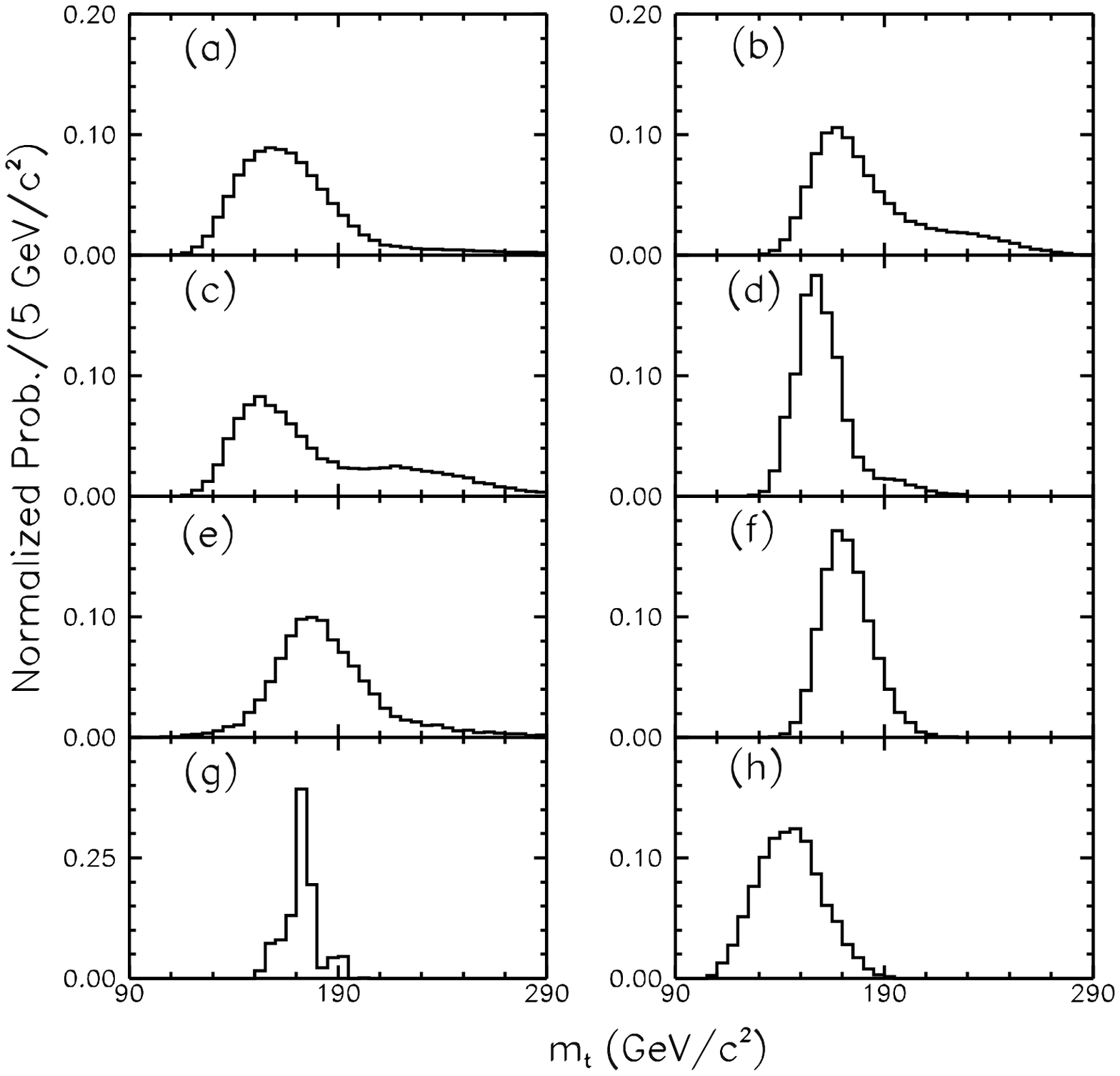}
\hskip 1.0 cm  \epsfxsize 2. truein \epsfbox{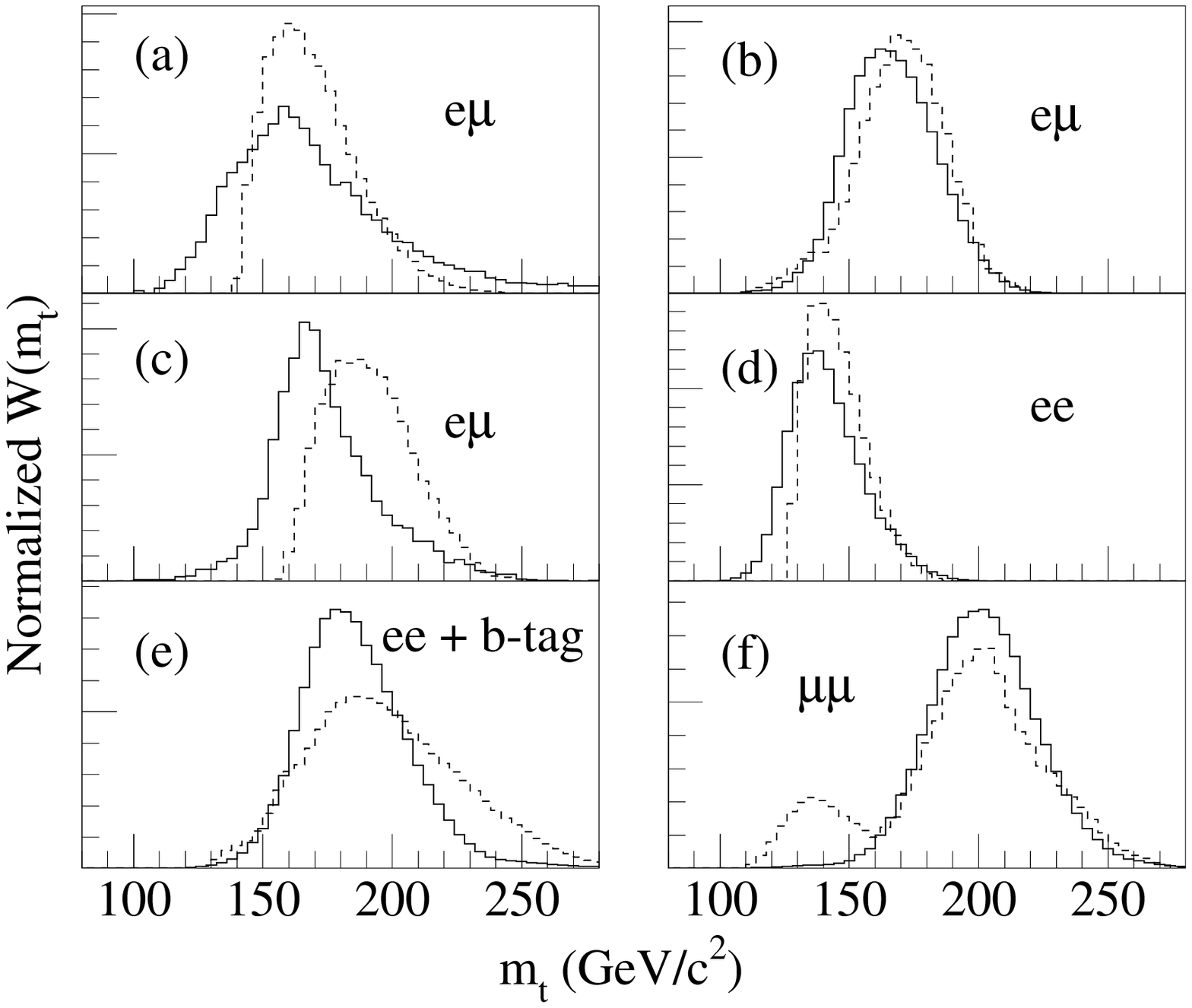}}
\vskip -.2 cm
\caption[]{
\label{dilw}
\small The normalized weight function for each dilepton candidate event 
 CDF(left) and D0(right).} 
\end{figure}

 For both weighting methods, the weights are summed over each lepton-jet 
paring, and over the detector resolution for jets and leptons by 
sampling ($\it{i.e.}$ fluctuating) the 
measured quantities many times according to their detector resolutions. 
The resulting weight functions for each CDF and D0 dilepton events, normalized
to unity,  are shown in Figure~\ref{dilw}. 

\begin{figure}[ht]	
\centerline{\epsfxsize 2.0 truein 
\epsfbox[65 130 554 641]{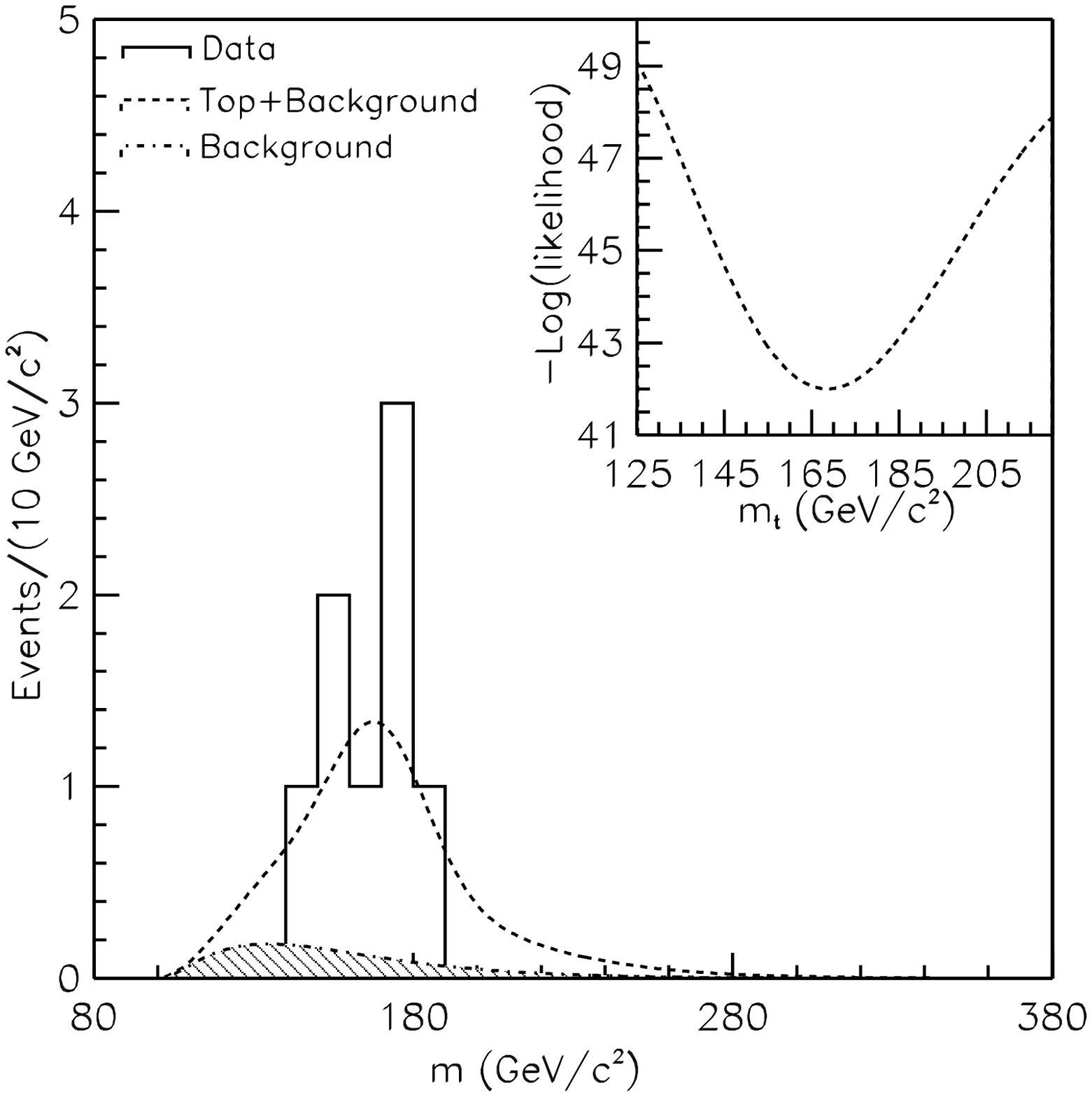 }
\hskip 1.0 cm \epsfxsize 2.0 truein \epsfbox{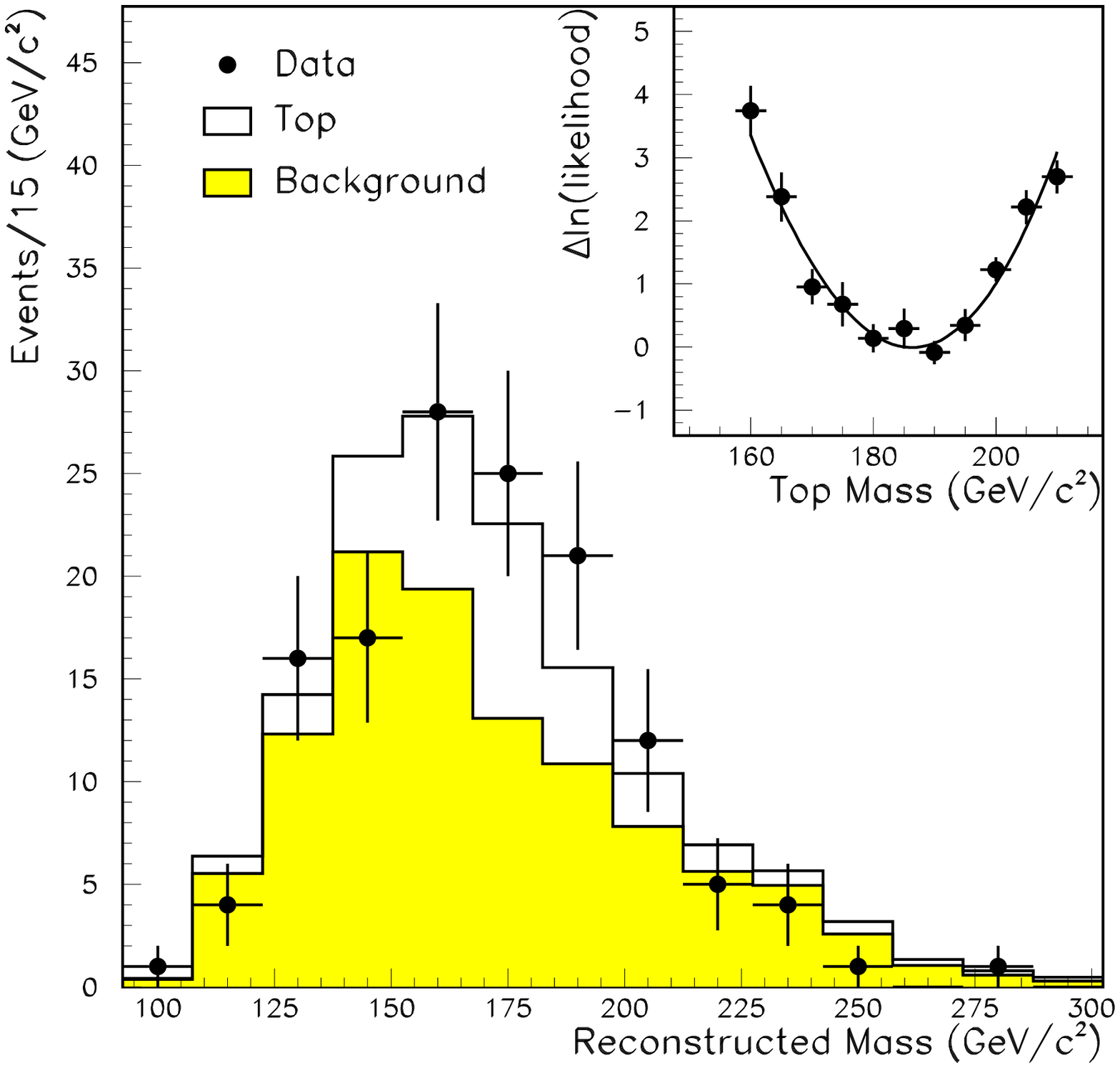}}
\vskip -.2 cm
\caption[]{
\label{cdfdilm}
\small Left: CDF reconstructed top mass for the eight dilepton events (solid)
with background(shaded) and fitted top(dashed). 
Right: CDF reconstructed top mass
in all hadronic events (point) with background(shaded) and fitted top (solid).
The inset shows the dependence of the likelihood on $m_t$.} 
\end{figure}

CDF assign each event a mass by averaging the two mass values 
corresponding to the weight closest to and greater than the half 
maximum weight on either side.  Figure~\ref{cdfdilm} shows the distribution of 
these masses, together with the Monte Carlo expectation for 
background and fitted top, from which CDF 
determines a top quark mass value of $167.4 \pm 10.3(stat.) \pm 4.8(syst.)$
GeV/c$^2$.

D0 accounts for the shape of the weight function by integrating 
the weights in five equally spaced mass bins and forming a four-dimensional
vector from the normalized weights. The top quark mass is extracted by 
comparing the distribution of observed events in this four-dimensional
vector space with the predicted densities for top signal and background. 
The results of these fits give a top quark mass value of 
 $168.4 \pm 12.3(stat.)\pm 3.6(syst.)$ GeV/c$^2$.

\subsection{All-hadronic Channel}
In this analysis~\cite{cdfaj} 
CDF selects $t\bar t$ events in which both $W$ bosons decay 
into quark-antiquark pairs, leading to an all hadronic final state. The study
of this channel, with a branching ratio of about 4/9, complements the 
leptonic modes, and the mass measurement takes advantage of  fully 
reconstructed final state, but suffers a very large QCD multijet background. 
To reduce this background, events with at least one identified  SVX $b$-jet 
are required to pass strict kinematic criteria that favor $t\bar t$ production
and decay. 

The data sample consist of 136 events, of which $108\pm 9$ events
are expected to come from background. 
Events are then reconstructed to the 
$t\bar t\rightarrow W^+b W^- \bar b$ hypothesis, where both $W$ bosons decay 
into a quark pair, with each quark associated to one of the six highest $\et$
jets. All the combinations are tried and the combination with lowest 
$\chi^2 < 10$ is chosen. 
The reconstructed 3-jet mass distribution 
is shown  in Figure~\ref{cdfdilm}, from which CDF measures
a top quark mass of $186.0 \pm 10.0(stat.) \pm 5.7(syst.)$ GeV/c$^2$. The 
overall systematic error has been revised from 
12.0 to 5.7 GeV/c$^2$~\cite{cdfdiln}. 

\subsection{A Combined Top Mass from CDF and D0} 

 The individual results from both CDF and D0 are combined according to 
a procedure developed by a joint CDF-D0 working group. The combined top mass 
from the Tevatron is $174.3\pm 3.2(stat)\pm 4.0 (sys)$ GeV/c$^2$. The good 
agreement among the measurements is reflected by a 
$\chi^2$ probability of 75\% of the mass average. Combining 
the statistical and systematic errors in quadrature, we find $m_t = 174.3\pm
5.1$ GeV/c$^2$, which gives the best measured quark mass 
($\delta m_t/m_t<3\%$). 

\subsection{Future Improvements at Run II} 

 In Run-II with 2 or more fb$^{-1}$ of integrated luminosity and the 
increased capabilities of the upgraded CDF and D0 detectors, we expect 
more than 1000 single tagged and about 600 double tagged $t\bar t$ events for
each experiment. It will allow us to measure $m_t$ down to statistical 
error of 0.5 GeV/c$^2$ by scaling the present results. The dominant 
systematic errors are expected to be the uncertainties in modeling gluon 
radiation and the jet energy scale of the detectors, which hopefully can be 
reduced to 2 GeV/c$^2$ by studying $W\rightarrow j_1 j_2$  
in the double tagged $t\bar t$ events and 32K $Z\rightarrow b\bar b$ events
from secondary vertex trigger.

\section{Acknowledgments} 

The author is grateful to all members of the CDF and D0 
Top and Exotic Physics groups for providing material in this paper. Especially, 
I would like to thank for: E. Barberis, J. Conway, R. Demina, L. Galtieri, 
J. Lys, M. Shapiro, F. Stichelbaut and D. Wood for their useful comments.

\end{document}